\documentclass[aip,apl,reprint,amsmath,amssymb]{revtex4-1}
\pdfoutput=1
\usepackage{graphicx}%
\usepackage{dcolumn}%
\usepackage{bm}%

\begin{document}

\title{Heteroepitaxy of Group IV-VI Nitrides by Atomic Layer Deposition}

\author{Jeffrey A. Klug}
\email{jklug@anl.gov}
\affiliation{Materials Science Division, Argonne National Laboratory, Argonne, Illinois 60439, USA}
\author{Nicholas G. Becker}
\affiliation{Materials Science Division, Argonne National Laboratory, Argonne, Illinois 60439, USA}
\affiliation{Department of Physics, Illinois Institute of Technology, Chicago, Illinois 60616, USA}
\author{Nickolas R. Groll}
\affiliation{Materials Science Division, Argonne National Laboratory, Argonne, Illinois 60439, USA}
\author{Chaoyue Cao}
\affiliation{Materials Science Division, Argonne National Laboratory, Argonne, Illinois 60439, USA}
\affiliation{Department of Physics, Illinois Institute of Technology, Chicago, Illinois 60616, USA}
\author{Matthew S. Weimer}
\affiliation{Materials Science Division, Argonne National Laboratory, Argonne, Illinois 60439, USA}
\affiliation{Department of Chemistry, Illinois Institute of Technology, Chicago, Illinois 60616, USA}
\author{Michael J. Pellin}
\affiliation{Materials Science Division, Argonne National Laboratory, Argonne, Illinois 60439, USA}
\author{John F. Zasadzinski}
\affiliation{Materials Science Division, Argonne National Laboratory, Argonne, Illinois 60439, USA}
\affiliation{Department of Physics, Illinois Institute of Technology, Chicago, Illinois 60616, USA}
\author{Thomas Proslier}
\email{prolier@anl.gov}
\affiliation{Materials Science Division, Argonne National Laboratory, Argonne, Illinois 60439, USA}


\begin{abstract}
Heteroepitaxial growth of selected group IV-VI nitrides on various orientations of sapphire ($\alpha$-Al$_2$O$_3$) 
is demonstrated using atomic layer deposition. High quality, epitaxial films are produced at significantly lower 
temperatures than required by conventional deposition methods. Characterization of electrical and superconducting 
properties of epitaxial films reveals a reduced room temperature resistivity and increased residual resistance ratio 
(RRR) for films deposited on sapphire compared to polycrystalline samples deposited concurrently on fused quartz 
substrates.
\end{abstract}
\maketitle


Transition metal nitrides exhibit a host of rich physics and functionality ranging from superconductivity 
to applications as refractory coatings.\cite{toth_1971} There is significant interest in epitaxial nitrides since 
epitaxy-induced changes in the crystallinity and morphology of thin films can influence both fundamental materials 
properties and device performance. Heteroepitaxial growth of MoN, NbN, and TiN has been explored by 
several methods, including reactive sputtering,\cite{noskov_1980_en,lloyd_nbn_2001,espiau_de_lamaestre_microstructure_2007} 
pulsed laser deposition (PLD),\cite{talyansky_tin_1998} chemical vapor deposition (CVD),\cite{wang_nbn_cvd_1996} and 
polymer assisted deposition (PAD).\cite{zhang_mon_pad_2011,*luo_mon_pad_2011,zou_nbn_pad_2008} Atomic layer deposition 
(ALD), which utilizes sequential self-limiting surface chemical reactions to deposit material in a layer-by-layer 
mode,\cite{ritala_book_2001, puurunen_jap_2005} offers several advantages over traditional growth techniques. ALD 
provides atomic-scale uniformity over large areas, unmatched conformality over complex-shaped substrates, and 
deposition temperatures well below those typically required by other methods. 

While the ALD of metal nitrides is known to result in the formation of crystalline films in many 
cases,\cite{miikkulainen_crystallinity_2013} little attention has been devoted to the investigation of heteroepitaxial 
growth on suitable single crystal substrates. Several groups have reported epitaxial ALD of 
AlN,\cite{khan_aln_1992} GaN,\cite{khan_sale_1992,*khan_sensors_1992,*khan_super_1993,*karam_gan_1995}, 
InN,\cite{nepal_epitaxial_2013} and In$_{x}$Ga$_{1-x}$N\cite{bedair_ingan_1997} films on sapphire 
($\alpha$-Al$_2$O$_3$) from organometallic precursors. Epitaxial GaN has also been reported on (111)GaAs using 
a chloride-ammonia process.\cite{tsuchiya_gan_1996} However, apart from the group III nitrides, epitaxy of nitrides 
has been entirely unexplored in the ALD literature despite a large volume of studies concerning transition metal nitride 
ALD.\cite{miikkulainen_crystallinity_2013,kim_review_2003} In this letter we report the ALD of 
epitaxial MoN, NbN, Nb$_{x}$Ti$_{1-x}$N, and TiN films on \textit{c}-, \textit{m}-, \textit{a}-, and \textit{r}-plane 
sapphire. Heteroepitaxial orientations were determined by x-ray diffraction (XRD) and the effects of substrate induced 
epitaxy on film resistivity and superconductivity are discussed. While several studies have examined superconductivity 
in ALD films, \cite{proslier_nbsi_2011,*klug_nbc_2011,*proslier_ecs_2011,*driessen_prl_2012,*coumou_tin_2013} the effects 
of crystallinity on film properties have thusfar not been reported.


ALD film growth was carried out in a custom-built hot-walled viscous flow reactor similar to that described 
elsewhere.\cite{elam_viscous_2002} During ALD, a 360 sccm flow of UHP N$_2$ at 1.0 Torr served as purge and carrier gas.
Prior to ALD, all substrates were cleaned via sonication in acetone, isopropanol, and deionized water. Nitride films 
were deposited at 450$~^{\circ}$C using metal chlorides (TiCl$_4$, NbCl$_5$, and MoCl$_5$) and ammonia 
(NH$_3$) as precursors. Nb$_x$Ti$_{1-x}$N samples were deposited using alternating TiN and NbN ALD cycles to achieve a 
desired composition. For these samples metallic Zn was used as an additional reactant for the reduction of the metal 
chlorides.\cite{ritala_zinc_1997} X-ray diffraction (XRD) and x-ray reflectivity (XRR) were performed with a Philips X'Pert MRD 
diffractometer using Cu K$\alpha$ radiation ($\lambda = 1.5418$~\AA) and operated at 30 kV / 30 mA. Incident x-ray 
beam conditioning was provided by a 60 mm graded parabolic W/Si mirror with a 0.8$^{\circ}$ acceptance angle and a 
1/32$^{\circ}$ divergence slit. The reflected beam was collected with a sealed proportional detector 
positioned behind a 0.27$^{\circ}$ parallel plate collimator and a pyrolytic graphite monochromator. Electrical 
resistivity down to 1.3 K was obtained via a four terminal measurement in a custom-built apparatus.


The epitaxial orientations determined by XRD for a series of MoN, NbN, TiN, and Nb$_{0.8}$Ti$_{0.2}$N samples are 
summarized in Table~\ref{tab:orient}. For clarity, planes and directions for MoN and Al$_2$O$_3$ are described using 
the four-axis hexagonal reference basis.\cite{otte_hex_1965} In all cases, films grown on sapphire were epitaxial while 
those grown concurrently on (001)Si and fused quartz (G.E. 124) substrates were polycrystalline. At least two twin 
variants were detected for all samples except the cases of MoN grown on \textit{c}- or \textit{m}-plane Al$_2$O$_3$. 

\begin{figure}
	\centering
		\includegraphics{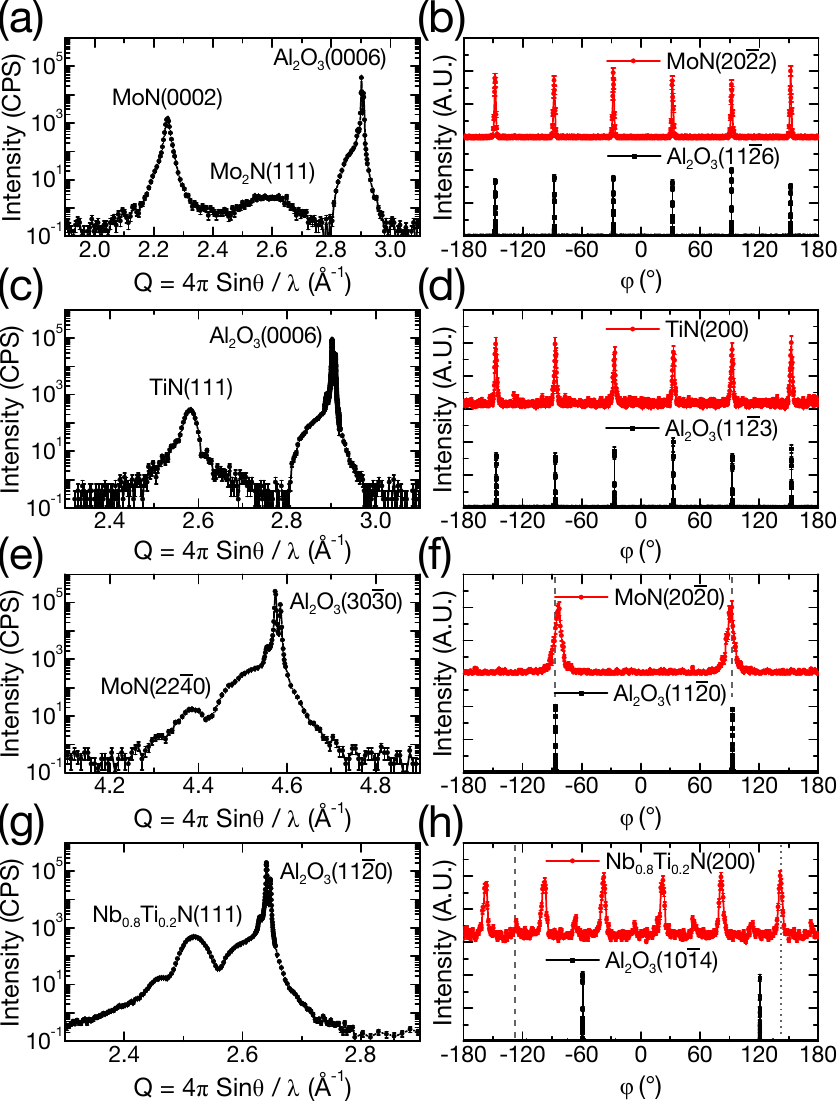}
		\caption{(Color online) Scattered x-ray intensity from [(a), (c), (e), and (g)] longitudinal scans along the specular 
		crystal truncation rod and [(b), (d), (f), and (h)] azimuthal scans through a set of off-specular reflections 
		demonstrate the OP and IP epitaxy, respectively, for [(a) and (b)] 80 nm MoN on \textit{c}-Al$_2$O$_3$, 
		[(c) and (d)] 20 nm TiN on \textit{c}-Al$_2$O$_3$, [(e) and (f)] 50 nm MoN on \textit{m}-Al$_2$O$_3$, and [(g) and 
		(h)] 15 nm Nb$_{0.8}$Ti$_{0.2}$N on \textit{a}-Al$_2$O$_3$. The dashed lines in (f) are a guide for the eye. The 
		dashed and dotted lines in (h) represent the $\phi$ positions of the [$\bar{1}100$] and [0001] directions, respectively, 
		of the Al$_2$O$_3$ substrate.}
		\label{fig:xrd}
\end{figure}

Specular and off-specular XRD scans of an 80 nm thick MoN film deposited on \textit{c}-Al$_2$O$_3$ are shown in 
Fig.~\ref{fig:xrd}(a) and~\ref{fig:xrd}(b), respectively. A single $\delta$-MoN orientation was observed,  
(0001)$\left\langle10\bar{1}0\right\rangle$MoN$\|$(0001)$\left\langle11\bar{2}0\right\rangle$Al$_2$O$_3$, which is 
consistent with the results reported for MoN/\textit{c}-Al$_2$O$_3$ prepared at 900$^{\circ}$ by 
PAD.\cite{zhang_mon_pad_2011,*luo_mon_pad_2011} From analysis of the (0002) and \{20$\bar{2}$2\} film reflections, 
the out-of-plane (OP) mosaic and in-plane (IP) texture of the MoN film were $1.12\pm0.02^{\circ}$ and $1.8\pm0.03^{\circ}$, 
respectively (FWHM). A broad peak corresponding to the (111) reflection of cubic $\gamma$-Mo$_2$N was observed with 
approximately 1/120 the integrated intensity of the MoN(0002) peak. This is consistent with a thin $\gamma$-Mo$_2$N 
layer at the film-substrate interface as observed in polycrystalline films deposited on 
AlN/quartz.\cite{suppl} Roughness of the thicker MoN layer ($\sim2$~nm) prevented accurate determination 
of the Mo$_2$N layer thickness by x-ray reflectivity. However, the width of the (111) peak established a lower limit 
of $\sim4$~nm on the Mo$_2$N thickness. The FWHM of the $\gamma$-Mo$_2$N(111) x-ray rocking curve was $1.6\pm0.3^{\circ}$,\cite{suppl} 
indicating that the Mo-rich layer is highly-ordered, although the IP orientation of this thin layer was not investigated 
due to practical limitations of the x-ray source. The same epitaxial relationship was obtained when MoN was deposited on 
\textit{c}-Al$_2$O$_3$ with an ALD-grown 6 nm thick epitaxial (0001)AlN buffer layer.

\begin{table}
\scriptsize
\caption{\label{tab:orient}Orientational relationships between ALD nitride films and single crystal sapphire substrates 
determined by x-ray diffraction.}
	\begin{ruledtabular}
	\begin{tabular}{ccc}
	\footnotesize{Film}&\footnotesize{Substrate}&\footnotesize{Epitaxial relationship}\\\hline\\
	MoN&\textit{c}-Al$_2$O$_3$&(0001)$\left\langle10\bar{1}0\right\rangle$MoN$\|$(0001)$\left\langle11\bar{2}0\right\rangle$Al$_2$O$_3$\\\\
	NbN&\textit{c}-Al$_2$O$_3$&$(111)[1\bar{1}0]$NbN$\|(0001)[10\bar{1}0]$Al$_2$O$_3$\\
	 & &$(111)[\bar{1}10]$NbN$\|(0001)[10\bar{1}0]$Al$_2$O$_3$\\\\
	TiN&\textit{c}-Al$_2$O$_3$&$(111)[1\bar{1}0]$TiN$\|(0001)[10\bar{1}0]$Al$_2$O$_3$\\
	 & &$(111)[\bar{1}10]$TiN$\|(0001)[10\bar{1}0]$Al$_2$O$_3$\\\\
	MoN&\textit{m}-Al$_2$O$_3$&$(11\bar{2}0)[1\bar{1}00]$MoN$\|(10\bar{1}0)[1\bar{2}10]$Al$_2$O$_3$\\\\
	Nb$_{0.8}$Ti$_{0.2}$N&\textit{a}-Al$_2$O$_3$&$(111)[1\bar{1}0]$Nb$_{0.8}$Ti$_{0.2}$N$\|(11\bar{2}0)[1\bar{1}00]$Al$_2$O$_3$\\
	 & &$(111)[\bar{1}10]$Nb$_{0.8}$Ti$_{0.2}$N$\|(11\bar{2}0)[1\bar{1}00]$Al$_2$O$_3$\\
	 & &$(111)[1\bar{1}0]$Nb$_{0.8}$Ti$_{0.2}$N$\|(11\bar{2}0)[0001]$Al$_2$O$_3$\\
	 & &$(111)[\bar{1}10]$Nb$_{0.8}$Ti$_{0.2}$N$\|(11\bar{2}0)[0001]$Al$_2$O$_3$\\\\
	MoN&\textit{r}-Al$_2$O$_3$&$[1\bar{2}12][10\bar{1}0]$MoN$\|(1\bar{1}02)[\bar{1}101]$Al$_2$O$_3$\\
	 & &$[1\bar{2}12][\bar{1}010]$MoN$\|(1\bar{1}02)[\bar{1}101]$Al$_2$O$_3$\\\\
	TiN&\textit{r}-Al$_2$O$_3$&$(135)[1\bar{2}1]$TiN$\|(1\bar{1}02)[\bar{1}101]$Al$_2$O$_3$\\
	 & &$(\bar{1}\bar{3}\bar{5})[\bar{1}2\bar{1}]$TiN$\|(1\bar{1}02)[\bar{1}101]$Al$_2$O$_3$\\
	\end{tabular}
	\end{ruledtabular}
\end{table}


ALD of NbN and TiN on \textit{c}-Al$_2$O$_3$ resulted in cubic (111)-oriented films with two twin IP variants in 
agreement with previous studies of sputtered NbN\cite{lloyd_nbn_2001,noskov_1980_en} and PLD-grown TiN 
films.\cite{talyansky_tin_1998} The XRD results obtained for a 20 nm thick TiN film are shown in Figs.~\ref{fig:xrd}(c) 
and \ref{fig:xrd}(d). The six TiN\{200\} peaks observed in Fig.~\ref{fig:xrd}(d) indicate the presence of two twin 
domains related by a 180$^{\circ}$ rotation about the [111] axis. This leads to a doubling of the expected threefold 
rotational symmetry about the [111] axis and mirrors the symmetry of the Al$_2$O$_3$ basal plane. The IP epitaxial 
relationship can therefore be described as $\pm[1\bar{1}0]$TiN$\|[10\bar{1}0]$Al$_2$O$_3$. The FWHM of the TiN(111) 
x-ray rocking curve\cite{suppl} was $0.0169\pm0.0005^{\circ}$ which indicates a significantly higher degree of OP 
alignment compared to the MoN/\textit{c}-Al$_2$O$_3$ sample. The reduced OP mosaicity of the TiN film may be related to 
the lack of a secondary phase at the film-substrate interface. However, analysis of the \{200\} $\phi$ scan 
[Fig.~\ref{fig:xrd}(d)] indicates a broader IP texture $2.83\pm0.08^{\circ}$ (FWHM) compared to that of the 
MoN/\textit{c}-Al$_2$O$_3$ sample which is likely a consequence of the twinned TiN microstructure. Identical measurements 
of a 35 nm thick film of the isostructural NbN found an analogous result.\cite{suppl} The OP mosaic and IP texture of 
the NbN film were $0.034\pm0.001^{\circ}$ and $1.98\pm0.03^{\circ}$, respectively.

The XRD results obtained for a 50 nm thick MoN film grown on \textit{m}-Al$_2$O$_3$ are presented in 
Figs.~\ref{fig:xrd}(e) and~\ref{fig:xrd}(f). The film is untwinned $\delta$-MoN with a single orientation described 
by $(11\bar{2}0)[1\bar{1}00]$MoN$\|(10\bar{1}0)[1\bar{2}10]$Al$_2$O$_3$ and no evidence of a cubic phase. The 
relative orientations of MoN and Al$_2$O$_3$ are approximately equivalent to the \textit{c}-Al$_2$O$_3$ case, with the 
hexagonal basal planes of film and substrate roughly parallel and the unit cell basis vectors offset by $30^{\circ}$ about the principal 
\textit{c}-axis. However, Fig~\ref{fig:xrd}(f) reveals that the MoN peak centers are shifted along the azimuthal axis 
$\sim2.5^{\circ}$ in opposite directions relative to each corresponding Al$_2$O$_3$ reflection. This is consistent with a 
tilting of the MoN(11$\bar{2}$0) plane away from the Al$_2$O$_3$(10$\bar{1}$0) by $\sim5^{\circ}$ about the MoN[$\bar{1}100$] 
axis. A rocking curve measurement\cite{suppl} of the MoN(22$\bar{4}$0) reflection yielded a FWHM of $1.60\pm0.07^{\circ}$ indicating 
an OP mosaic comparable to the MoN/\textit{c}-Al$_2$O$_3$ sample. However, the azimuthal widths of the (20$\bar{2}$0) 
and (02$\bar{2}$0) MoN peaks were $8.0\pm0.6^{\circ}$, indicating a considerably wider variation in IP texture. While 
numerous factors can influence film microstructure, such as differing modes of interfacial strain relaxation, the broadened 
IP alignment is not surprising given the lower symmetry of the MoN(11$\bar{2}$0) and Al$_2$O$_3$(10$\bar{1}$0) planes compared 
to the high-symmetry (0001) planes.


On \textit{a}-Al$_2$O$_3$, ALD of a 15 nm thick film of Nb$_{0.8}$Ti$_{0.2}$N resulted in a cubic (111)-oriented film with 
two sets of twinned IP domains similar to the case of NbN/\textit{a}-Al$_2$O$_3$ reported in Ref.~\onlinecite{lloyd_nbn_2001}. 
Fig.~\ref{fig:xrd}(g) demonstrates OP orientation of the film, with only Nb$_{0.8}$Ti$_{0.2}$N(111) and Al$_2$O$_3$(11$\bar{2}$0) 
reflections observed in a specular XRD scan. In contrast, the azimuthal scans shown in \ref{fig:xrd}(h) revealed the presence 
of twelve equally-spaced Nb$_{0.8}$Ti$_{0.2}$N\{200\} peaks rather than the three expected from a single crystallographic domain. 
The twelve observed \{200\} peaks comprise two distinct sets which are offset by 30$^{\circ}$ ($modulo~60$) and differ in 
integrated intensity by a factor of $\sim6$. This indicates that the film contains primary and secondary pairs of 180$^{\circ}$ 
twin IP domains which are offset by 90$^{\circ}$ about the [111] axis. The IP orientations of the primary and secondary sets of 
twins were determined by comparison with a $\phi$ scan through the Al$_2$O$_3$(10$\bar{1}$4) and (01$\bar{1}\bar{4}$) reflections 
[Fig.~\ref{fig:xrd}(h)]. The orthogonal IP substrate directions $[1\bar{1}00]$ and [0001] are oriented $\Delta\phi=-68.454^{\circ}$ 
and $\Delta\phi=21.546^{\circ}$, respectively, from the projection of the (10$\bar{1}$4) in the Al$_2$O$_3$(11$\bar{2}$0) plane. 
Since the $\left\langle1\bar{1}0\right\rangle$ directions are offset 30$^{\circ}$ from the projection of the 
$\left\langle200\right\rangle$ in the (111) plane, the IP orientations of the primary and secondary twins are therefore described 
by $\pm[1\bar{1}0]$Nb$_{0.8}$Ti$_{0.2}$N$\|[1\bar{1}00]$Al$_2$O$_3$ and $\pm[1\bar{1}0]$Nb$_{0.8}$Ti$_{0.2}$N$\|[0001]$Al$_2$O$_3$, 
respectively. The OP mosaic and IP texture of the Nb$_{0.8}$Ti$_{0.2}$N film were $0.0186\pm0.0005^{\circ}$ and $4.6\pm0.2^{\circ}$.

\begin{figure}[t!]
	\centering
		\includegraphics{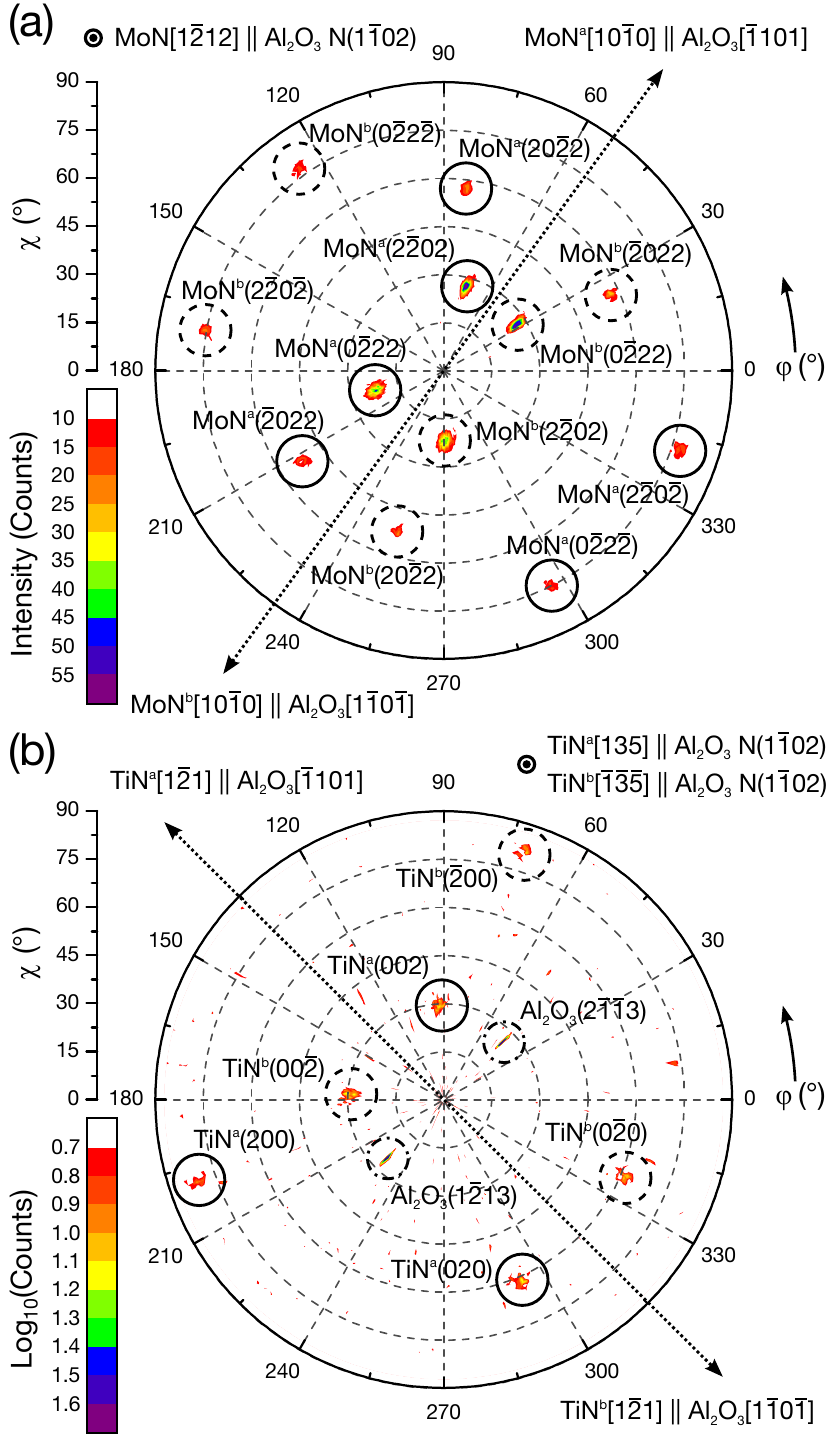}
		\caption{(Color online) Pole figures of the (a) hexagonal \{20$\bar{2}$2\} Bragg reflections for an 80 nm MoN film 
		and (b) the cubic \{200\} Bragg reflections for a 20 nm TiN film deposited on \textit{r}-plane sapphire. Peaks 
		from each twin variant are labeled with superscripts \textit{a} and \textit{b} and marked with solid and dashed circles. 
		In the TiN measurement, peaks from the Al$_2$O$_3$ substrate are marked with dash-dot circles.}
		\label{fig:tinpole}
\end{figure}


On \textit{r}-Al$_2$O$_3$, the growth directions of both hexagonal MoN and cubic TiN are such that no allowed reflections 
are positioned along the specular Al$_2$O$_3$(1$\bar{1}$02) rod. Consequently, x-ray pole figures of the MoN\{20$\bar{2}$2\} 
and TiN\{200\} reflections were measured [Figs.~\ref{fig:tinpole}(a) and \ref{fig:tinpole}(b), respectively] to determine 
film orientation in each case. Twelve MoN\{20$\bar{2}$2\} peaks were observed for an 80 nm thick MoN/\textit{r}-Al$_2$O$_3$ 
film, as shown in Fig.~\ref{fig:tinpole}(a). Analysis of the data found that the film is [1$\bar{2}$12]-oriented, with two 
IP twin variants related by a 180$^{\circ}$ rotation about the OP MoN[1$\bar{2}$12] axis. For clarity, reflections from the 
two twins are labeled MoN$^{a}$ and MoN$^{b}$ in Fig.~\ref{fig:tinpole}(a). Comparison with a $\phi$ scan through the 
Al$_2$O$_3$(2$\bar{1}\bar{1}$3) and (1$\bar{2}$13) reflections (not shown) was used to determine the IP epitaxial orientation 
of the film, which is described by $\pm[10\bar{1}0]$MoN$\|[\bar{1}101]$Al$_2$O$_3$. Small displacements of the MoN peaks from 
their symmetric postions indicate that the [1$\bar{2}$12] directions of the film domains are tilted from 
Al$_2$O$_3$(1$\bar{1}$02) by $\sim3^{\circ}$. 

The TiN\{200\} x-ray pole figure obtained for a 20 nm TiN/\textit{r}-Al$_2$O$_3$ sample is shown in Fig.~\ref{fig:tinpole}(b). 
Six film peaks were observed indicating the presence of two twin variants 
$\pm(135)[1\bar{2}1]$TiN$\|(1\bar{1}02)[\bar{1}101]$Al$_2$O$_3$ related by a 180$^{\circ}$ rotation about the IP 
TiN[$\bar{13}.\bar{4}5$] axis (or, equivalently, 180$^{\circ}$ rotations about both the OP [135] and IP [1$\bar{2}$1] axes). In Fig.~\ref{fig:tinpole}(b), peaks from the (135)- and ($\bar{1}\bar{3}\bar{5}$)-oriented domains are labeled TiN$^{a}$ and TiN$^{b}$, respectively. The observed structure was equivalent to that reported for sputtered NbN on 
\textit{r}-Al$_2$O$_3$.\cite{noskov_1980_en,espiau_de_lamaestre_microstructure_2007} The $(2\bar{1}\bar{1}3)$ and $(1\bar{2}13)$ 
peaks from the Al$_2$O$_3$ substrate were visible in the TiN\{200\} pole figure due to the small difference in d-spacing between 
the TiN and Al$_2$O$_3$ reflections (2.1209 \AA~ and 2.0853 \AA, respectively), and conveniently illustrate the orientational 
registry of the film and substrate.

\begin{figure}[t!]
	\centering
		\includegraphics{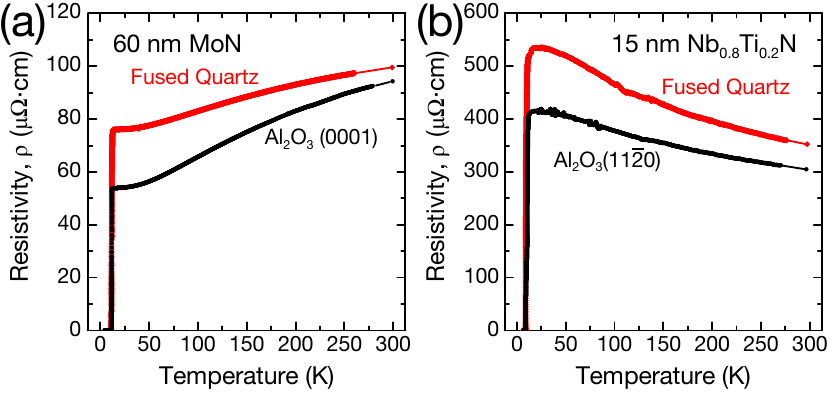}
		\caption{(Color online) Electrical resistivity versus temperature measured for (a) a 60 nm MoN film deposited concurrently 
		on (red diamonds) fused quartz and (black circles) \textit{c}-Al$_2$O$_3$, and (b) a 15 nm Nb$_{0.8}$Ti$_{0.2}$N film deposited 
		concurrently on (red diamonds) fused quartz and (black circles) \textit{a}-Al$_2$O$_3$.}
		\label{fig:rhovst}
\end{figure}

A comparison of electrical transport measurements of epitaxial and polycrystalline films found consistently more metallic 
behavior above the superconducting $T_{\rm c}$ in films deposited on sapphire with relatively minor change in $T_{\rm c}$. 
Figs.~\ref{fig:rhovst}(a) and \ref{fig:rhovst}(b) show the film resistivity, $\rho(T)$, measured for two different cases, 
respectively: a relatively thick 60 nm MoN film deposited on fused quartz and \textit{c}-Al$_2$O$_3$ and a thin 15 nm 
Nb$_{0.8}$Ti$_{0.2}$N film deposited on quartz and \textit{a}-Al$_2$O$_3$. In both cases, the critical temperature, 
$T_{\rm c0}$, of the film on sapphire was $<2$\% higher than the film on fused quartz. However, for 60 nm MoN, the room temperature 
resistivity, $\rho_{300\rm K}$ is $\sim5$\% lower and the residual resistance ratio (RRR) ($\rho_{300\rm K}/\rho_{20\rm K}$) 
is $\sim30$\% higher for the film grown on sapphire compared to the film on fused quartz. Likewise, for 15 nm Nb$_{0.8}$Ti$_{0.2}$N, 
$\rho_{300\rm K}$ and RRR for the sapphire sample were $\sim15$\% lower and $\sim10$\% higher, respectively, compared to the 
quartz sample. A similar effect was observed in Ref.~\onlinecite{espiau_de_lamaestre_microstructure_2007} where a comparison 
of ultrathin twinned and untwinned NbN films found that while the presence of twinning resulted in 15\% $\rho_{300\rm K}$ and 
a 20\% lower RRR ($\rho_{300\rm K}/\rho_{20\rm K}$), the $T_{\rm c}$ was reduced by $<2$\%.

In summary, high quality epitaxial films of MoN, NbN, TiN, and Nb$_x$Ti$_{1-x}$N were prepared on sapphire substrates by ALD 
at a low growth temperature of 450$~^{\circ}$C. Observed orientations of MoN, NbN, and TiN deposited on \textit{c}-Al$_2$O$_3$, 
and Nb$_x$Ti$_{1-x}$N on \textit{a}-Al$_2$O$_3$ were consistent with prior literature. Untwinned (11$\bar{2}$0)MoN was grown on 
\textit{m}-Al$_2$O$_3$, while on \textit{r}-Al$_2$O$_3$ both (135)TiN and [1$\bar{2}$12]MoN grow with two IP twin domains. 
Epitaxial films were found to be more metallic than polycrystalline samples deposited concurrently on fused quartz. These results 
demonstrate the utility of ALD for the synthesis of epitaxial films of transition metal nitrides. This has significant implications 
for applications of nitride thin films where high crystalline quality is required in combination with moderate growth temperature, 
large-scale uniformity, precise thickness control, or conformal coating over complex-shaped surfaces.


This work was supported by the U.S. Department of Energy, Office of Science under contract No. DE-AC02-06CH11357. We thank 
J. D. Emery for his critical reading of our manuscript.

\bibliography{Klug_EpitaxialNitrides_arxiv}

\begin{thebibliography}{30}
\bibitem [{sup()}]{suppl}%
  \BibitemOpen
  \href@noop {} {}\bibinfo {note} {See supplimentary material at [url will be
  inserted by AIP] for additional XRD data.}\BibitemShut
  {Stop}%
\end{thebibliography}

\end{document}